\documentclass[11pt]{article}
\usepackage{amsmath,amsfonts,latexsym,graphicx,amssymb}
\date{}
\def\la{\langle\,}
\def\r{\,\rangle}

\newcommand{\bbC}{\mathbb{C}}

\def\oper{{\mathchoice{\rm 1\mskip-4mu l}{\rm 1\mskip-4mu l}%
{\rm 1\mskip-4.5mu l}{\rm 1\mskip-5mu l}}}
\def\con{{}_{\_\rule{-1pt}{0pt}\_}
\rule{-2pt}{0pt}\raise1.5pt\hbox{$\mid$}\hspace{2pt}}

\title{\bf Spectral properties of the squeeze operator}
\author{Dariusz Chru\'sci\'nski\\
 Institute of Physics, Nicolaus Copernicus University\\
 ul. Grudzi\c{a}dzka 5/7, 87-100 Toru\'n, Poland}

\begin{document}

\maketitle

\begin{abstract}

We show that a single-mode squeeze operator $S(z)$ being an
unitary operator with a purely continuous spectrum gives rise to a
family of discrete real generalized eigenvalues. These eigenvalues
are closely related to the spectral properties of $S(z)$ and the
corresponding generalized eigenvectors may be interpreted as
resonant states well known in the scattering theory. It turns out
that these states entirely characterize the action of $S(z)$. This
result is then generalized to $N$-mode squeezing.

\end{abstract}

%\vspace{.7cm}

%\noindent {\bf Mathematical Subject Classifications (2000):}
%46E10, 46F05, 46N50, 47A10.

%\vspace{.3cm}

%\noindent {\bf Key words:} quantum mechanics, distributions,
%spectral theorem, Gelfand triplets.

%\newpage

\section{Introduction}
\setcounter{equation}{0}

\numberwithin{equation}{section}

Squeezed states play a prominent role in the modern quantum
optics, see e.g.  \cite{Walls,book1,book2}. They are quantum
states for which no classical analog exists. Recently, they have
drown a great deal of interest in connection with quantum
teleportation. Squeezed states of light were successfully
teleported in the experiment reported by the group of Furusawa
\cite{Furusawa}. Mathematical properties of these states were
investigated in a series of papers
\cite{Stoler,Yuen,Hollenhorst,Fisher,Truax}. In the present paper
we analyze the spectral properties of single-mode $S(z)$ and
$N$-mode $S_N(\widehat{Z})$ squeeze operators. Clearly, squeeze
operators are unitary and hence their spectra define a subset of
complex numbers with modulus one. It is easy to show that the
spectrum of the squeeze operator operator is purely continuous and
cover the entire unit circle on the complex plane. However, it is
not the whole story. Actually, it is easy to note that $S(z)$
displays two families of discrete real eigenvalues. Clearly, these
eigenvalues are not proper, that is, the corresponding
eigenvectors do not belong to the corresponding Hilbert space of
square integrable functions. One of these families was reported in
a series of papers by Jannussis et al. \cite{series1,series2}.
However, the interpretation of this result was not clear (it was
criticized by Ma et al. \cite{Ma} who stressed  that proper
eigenvalues of the squeeze operator do not exist). In the present
paper we are going to clarify this problem. In particular we show
that $S(z)$ being an unitary operator with a purely continuous
spectrum does indeed possesses discrete eigenvalues which are
closely related to its spectral properties. The corresponding
eigenvectors may be interpreted as resonant states well known in
the scattering theory \cite{scattering,Gamov}. We show that
restricting to a suitable class of states (e.g. coherent states do
belong to this class) the action of $S(z)$ may be entirely
characterized in terms  of these discrete eigenvalues and the
corresponding eigenvectors. This observation is then generalized
to two-mode squeezing and finally to $N$-mode squeezing.

\section{Single-mode squeezing}

A single-mode squeeze operator is defined by
\begin{equation}\label{S}
  S(z) = \exp\left( \frac 12 \Big[ za^{\dag 2} - z^*
  a^2 \Big] \right) \ ,
\end{equation}
where $z$ is a complex number and $a$ ($a^\dag)$ is the photon
annihilation (creation) operator which obey the standard
commutation relation $[a,a^\dag]=1$. Clearly, $S(z)$ may be
represented as $S(z)=\exp(iH(z))$, with
\begin{equation}\label{}
  H(z) = \frac{1}{2i}\ \Big( za^{\dag 2} - z^*
  a^2 \Big)\ .
\end{equation}
Now, to investigate spectral properties of $H(z)$ let us consider
a unitarily equivalent operator $R^\dag(\varphi) H(z) R(\varphi)$,
where $R(\varphi)$ is a single-mode rotation
\cite{Schumaker1,Schumaker2,Schumaker3}
\begin{equation}\label{R}
  R(\varphi) = \exp(i\varphi\, a^\dag a) \ .
\end{equation}
One has
\begin{equation}\label{}
R^\dag(\varphi) H(z) R(\varphi) = H(ze^{-2i\varphi})\ ,
\end{equation}
and hence, for  $\varphi= \theta/2$, where $z = r e^{i\theta}$, it
shows that $H(z)$ is unitarily equivalent to $H(r)$. Introducing
two quadratures $x$ and $p$ {\em via}
\begin{equation}\label{}
a = \frac{x+ip}{\sqrt{2}}\ , \ \ \ \ a^\dag =
\frac{x-ip}{\sqrt{2}}\ ,
\end{equation}
one finds the following formula for $H(r)$:
\begin{equation}\label{}
  H(r) = - \frac r2 \, (xp + px) \ .
\end{equation}
 Spectral properties of $H(r)$ were
recently investigated in \cite{D1} (see also \cite{Bollini}) in
connection with quantum dissipation. Note, that the classical
Hamilton equations implied by the Hamiltonian $H=-rxp$:
\begin{equation}\label{}
  \dot{x} = - rx\ , \ \ \ \ \ \dot{p} = rp\ ,
\end{equation}
describe the damping of $x$ and pumping of $p$. This is a
classical picture of the squeezing process.
 It turns out that $H(r)$ has
purely continuous spectrum covering the whole real line. Hence, it
is clear that it does not have any proper eigenvalue. Using
standard Schr\"odinger representation for $x$ and $p= - id/dx$,
the corresponding eigen-problem $H(r)\psi = E\psi$ may be
rewritten as follows
\begin{eqnarray}  \label{eq}
x\frac{d}{dx}\psi(x) = - \left( i\frac Er  + \frac 12 \right)
\psi(x) \ .
\end{eqnarray}
 Note,
that $H(r)$ is parity invariant and hence each generalized
eigenvalue $E\in \mathbb{R}$ is doubly degenerated. Therefore, two
independent solutions of (\ref{eq}) are given by
\begin{eqnarray} \label{psi-E}
\psi^E_\pm(x) = \frac{1}{\sqrt{2\pi r}} \, x^{-(iE/r + 1/2)}_\pm\
,
\end{eqnarray}
where $x^\lambda_\pm$ are distributions defined as follows
\cite{Gelfand-S} (see also \cite{Kanwal}):
\begin{eqnarray} x^\lambda_+ := \left\{
\begin{array}{ll} x^\lambda & \ \ \ x\geq 0 \\ 0 &\ \ \ x<0
\end{array} \right. \ , \hspace{1cm} x^\lambda_- := \left\{
\begin{array}{cl} 0 & \ \ \  x\geq 0 \\ |x|^\lambda  &\ \ \ x<0
\end{array} \right. \ , \end{eqnarray} with $\lambda \in \bbC$ (basic
properties of $x^\lambda_\pm$ are collected in \cite{D1}). These
generalized eigenvectors $\psi^E_\pm$ are complete
\begin{eqnarray}
\int \overline{\psi^{E}_\pm(x)}\psi^{E}_\pm(x')\, dE =
\delta(x-x') \ ,
\end{eqnarray}
and $\delta$-normalized
\begin{eqnarray}
\int \overline{\psi^{E_1}_\pm(x)}\psi^{E_2}_\pm(x)\, dx =
\delta(E_1-E_2) \ .
\end{eqnarray}
Hence they give rise to the following spectral resolution of
$H(r)$:
\begin{eqnarray} \label{GMH-1}
{H}(r) = \sum_\pm \int\, E\, | \psi^E_\pm \r \la \psi^E_\pm|\, dE
\ ,
\end{eqnarray}
and the corresponding spectral resolutions of squeeze operator
$S(r)$ immediately follows
\begin{eqnarray} \label{S-spectral}
S(r) =  \sum_\pm \int\, e^{iE}\, | \psi^E_\pm \r \la \psi^E_\pm|\,
dE \ .
%\nonumber \\  &=& \sum_\pm \int\,  e^{iE}\, |
%F[\psi^{-E}_\pm] \r \la F[\psi^{-E}_\pm]|\, dE \ .
\end{eqnarray}
Now, let us observe that $ FH=-HF$, where $F$ denotes the Fourier
transformation. Hence, if $H(r)\psi^E = E\psi^E$, then
$H(r)F[\psi^{-E}] = EF[\psi^{-E}]$. Therefore, the family
$F[\psi^{-E}_\pm]$ defines another system of complete and
$\delta$-normalized generalized eigenvectors of $H(r)$. Note that
the action of $S(r)$ is defined by
\begin{equation}\label{}
  S(r)\psi(x) =
  e^{-r/2}\psi(e^{-r}x)\ ,
\end{equation}
and its Fourier transform
\begin{equation}\label{}
F[S(r)\psi](p) =
  e^{r/2}F[\psi](e^{r}p)\ ,
\end{equation}
due to $FS(r) = S(-r)F$, that is, if the fluctuations of $p$ are
reduced then the fluctuation of $x$ are amplified and vice versa.

\section{Discrete real eigenvalues of $S(r)$}

Surprisingly, apart from the continuous spectrum $H(r)$ gives rise
to the following families of complex discrete eigenvalues
\cite{D1}
\begin{eqnarray}
H(r) f^\pm_n = \pm E_n \, f^\pm_n\ ,
\end{eqnarray}
where
\begin{eqnarray} \label{En}
E_n = ir \left( n + \frac 12 \right) \ ,
\end{eqnarray}
and
\begin{eqnarray}  \label{fn} f^-_n(x) = \frac{(-1)^n}{\sqrt{n!}}\,
\delta^{(n)}(x)\ , \hspace{1cm} f^+_n(x) = \frac{x^n}{\sqrt{n!}}\
. \end{eqnarray} Interestingly they satisfy the following
properties:
\begin{eqnarray} \label{P1}
\int_{-\infty}^\infty f^+_n(x)\, f^-_m(x)\, dx = \delta_{nm}\ ,
\end{eqnarray}
and
\begin{eqnarray}   \label{P2}
\sum_{n=0}^\infty\, f^+_n(x)\, f^-_n(x') = \delta(x-x')\ .
\end{eqnarray}
It implies that
\begin{equation}\label{}
S(r) f^\pm_n = e^{\pm iE_n} \, f^\pm_n\ ,
\end{equation}
which shows that $S(r)$ displays two families of purely real
generalized eigenvalues
\begin{equation}\label{}
  s^\pm_n = \exp\left[ \pm r\left(n + \frac 12 \right)\right]\ .
\end{equation}
A family $s^+_n$ was already derived by Jannussis et al., see e.g.
formula (2.3) in \cite{series2}, but they overlooked the second
one $s^-_n$. How to interpret these eigenvalues? It turns out that
one recover $E_n$ and $f^\pm_n$ by studying a continuation of
generalized eigenvectors $\psi^E_\pm$ and $F[\psi^{-E}_\pm]$ into
the energy complex plane $E \in \bbC$. Both $\psi^E$ and
$F[\psi^{-E}]$ display singular behavior when $E$ is complex:
$\psi^E_\pm$ has simple poles at $E=-E_n$, whereas
$F[\psi^{-E}_\pm]$ has simple poles at $E=+E_n$, with $E_n$
defined in (\ref{En}). Moreover, their residues  correspond, up to
numerical factors, to the eigenvectors $f^\pm_n$:
\begin{eqnarray} \mbox{Res}(\psi^E_\pm(x);-E_n) \ \sim \ f^-_n \ ,
\end{eqnarray} and
\begin{eqnarray} \mbox{Res}(F[\psi^{-E}_\pm(x)];+E_n) \ \sim \
f^+_n \ .
\end{eqnarray}
Such eigenvectors are well known in scattering theory as resonant
states, see e.g. \cite{scattering} and references therein. In the
so called rigged Hilbert space to quantum mechanics these states
are also called Gamov vectors \cite{Gamov}. To see the connection
with the scattering theory let us observe that under the following
canonical transformation:
\begin{equation}\label{}
  x = \frac{rQ - P}{\sqrt{2r}}\ , \hspace{1cm}  p = \frac{rQ + P}{\sqrt{2r}}\ ,
\end{equation}
$H(r)$ transforms into the unitarily equivalent operator \cite{D2}
\begin{equation}\label{}
  H(r) \ \longrightarrow \ H_{\rm io} = \frac 12 (P^2 - r^2Q^2)\ ,
\end{equation}
which represents the Hamiltonian of the called an inverted or
reversed oscillator (or equivalently a potential  barrier `$- r^2
Q^2/2$') and it was studied by several authors in various contexts
\cite{Wheeler,Friedman,Ann1,Ann2,Castagnino,Shimbori1}. An
inverted oscillator $H_{\rm io}$ corresponds to the harmonic
oscillator with a purely imaginary frequency $\omega = \pm ir$ and
hence the harmonic oscillator spectrum `$\omega(n + 1/2)$' implies
`$\pm ir(n+1/2)$' as generalized eigenvalues of $H_{\rm io}$.

\section{A new representation of $S(r)$}

Interestingly, the action of $S(r)$ may be entirely characterized
in terms of $f^\pm_n$ and $s^\pm_n$. Indeed, consider a space
$\cal D$ of smooth functions $\psi=\psi(x)$ with compact supports,
i.e. $\psi(x)=0$ for $|x|>a$ for some positive $a$ (depending upon
chosen $\psi$), see e.g. \cite{Yosida}. Clearly, $\cal D$ defines
a subspace of square integrable functions $L^2( \mathbb{R})$.
Moreover, let ${\cal Z} = F[{\cal D}]$, that is, $\psi \in {\cal
Z}$ if $\psi=F[\phi]$ for some $\phi \in {\cal D}$. It turns out
\cite{Yosida} that $\cal D$ and $\cal Z$ are isomorphic and ${\cal
D} \cap {\cal Z} = \emptyset$. Now, any function $\phi$ from $\cal
Z$ may be expanded into Taylor series and hence
\begin{eqnarray}
\phi(x) = \sum_{n=0}^\infty \frac{\phi^{(n)}(0)}{n!} x^n =
\sum_{n=0}^\infty f^+_n(x) \la f^-_n|\phi\r \ .
\end{eqnarray}
On the other hand, for any $\phi \in {\cal D}$, its Fourier
transform $F[\phi] \in {\cal Z}$, and
\begin{eqnarray}
\phi(x) &=& \frac{1}{\sqrt{2\pi}} \int e^{ikx} F[\phi](k) dk =
\frac{1}{\sqrt{2\pi}} \int e^{ikx} \sum_{n=0}^\infty
\frac{ F[\phi]^{(n)}(0)}{n!} k^n \, dk \nonumber \\
&=& \sum_{n=0}^\infty F[f^+_n](x) \la f^-_n|F[\phi]\r =
%\sum_{n=0}^\infty F[f^+_n](x) \la F[f^-_n]|\phi\r \nonumber \\ &=&
\sum_{n=0}^\infty f^-_n(x) \la f^+_n|\phi\r \ .
\end{eqnarray}
Hence, we have two  decompositions of the identity operator
\begin{eqnarray}  \label{Z} \oper =  \sum_{n=0}^\infty |f^+_n\r \la f^-_n|
\hspace{1cm} \mbox{on}\ \ \ {\cal Z} \ , \end{eqnarray} and
\begin{eqnarray} \label{D} \oper =  \sum_{n=0}^\infty |f^-_n\r \la
f^+_n| \hspace{1cm} \mbox{on}\ \ \ {\cal D}\ .
\end{eqnarray}
It implies the following representations of the squeeze operator
$S(r)$:
\begin{eqnarray}  \label{S-Z} S(r) =  \sum_{n=0}^\infty \, s^-_n\, |f^+_n\r \la f^-_n|
\hspace{1cm} \mbox{on}\ \ \ {\cal Z} \ , \end{eqnarray} and
\begin{eqnarray} \label{S-D} S(r) =  \sum_{n=0}^\infty \, s^+_n\, |f^-_n\r \la
f^+_n| \hspace{1cm} \mbox{on}\ \ \ {\cal D}\ .
\end{eqnarray}
It should be stressed that the above formulae for $S(r)$ are not
spectral decompositions and  they valid only on $\cal Z$ and $\cal
D$, respectively (its spectral decomposition is given in formula
(\ref{S-spectral})). Note, that $S(r)$ maps $\cal Z$ into $\cal D$
and using (\ref{S-Z}) one has
\begin{equation}\label{}
  S^\dag(r) = \sum_{n=0}^\infty \, s^-_n\, |f^-_n\r \la f^+_n| =
  S(-r) \hspace{1cm} \mbox{on}\ \ \ {\cal D}\ .
\end{equation}
Conversely, $S(r)$ represented by (\ref{S-D}) maps $\cal D$ into
$\cal Z$ and $S^\dag(r) = S(-r)$ on $\cal Z$. It shows that
squeezing of $x\ (p)$ corresponds to amplifying of $p\ (x)$.
Clearly, a general quantum state $\psi \in L^2( \mathbb{R})$
belongs neither to $\cal D$ nor to $\cal Z$. An example of quantum
states belonging to $\cal Z$ is a family of Glauber coherent
states $|\alpha\rangle$. Consider e.g. a squeezed vacuum
$S(r)\psi_0$, where $\psi_0(x) = \pi^{-1/4}e^{-x^2/2}$. One has
\begin{equation}\label{}
  \psi_0(x) = \frac{1}{\pi^{1/4}}\, \sum_{n=0}^\infty\,
  \frac{(-1)^n}{\sqrt{(2n)!}}\, f^+_{2n}(x)\ ,
\end{equation}
and
\begin{equation}\label{}
  S(r)\psi_0(x) = \frac{e^{-r/2}}{\pi^{1/4}}\,  \sum_{n=0}^\infty\,
  \frac{(-e^{-2r})^n}{\sqrt{(2n)!}}\, f^+_{2n}(x) =
\frac{e^{-r/2}}{\pi^{1/4}}\,   \sum_{n=0}^\infty\,
  \frac{(-1)^n}{\sqrt{(2n)!}}\, f^+_{2n}(e^{-r}x)\ .
\end{equation}
The similar formulae hold for $S(r)|\alpha\r$.

\section{Two-mode squeezing}

Consider now a two-mode squeeze operator
\cite{Schumaker1,Schumaker2,Schumaker3}
\begin{equation}\label{}
  S_2(z) = \exp\Big( z a^\dag_1 a^\dag_2 - z^* a_1 a_2 \Big)\ ,
\end{equation}
where $a^\dag_k$ and $a_k$ are creation and annihilation operators
for two modes $k=1,2$. Introducing 2-dimensional vectors
\begin{equation}\label{}
  \underline{a}^T = (a_1,a_2)\ ,\ \ \ \ \ \  (\underline{a}^\dag)^T = (a^\dag_1,a^\dag_2)\ ,
\end{equation}
one finds
\begin{equation}\label{}
  S_2(z) = \exp\Big( z (\underline{a}^\dag)^T \sigma_1 \underline{a}^\dag   - z^*
  \underline{a}^T \sigma_1 \underline{a} \Big)\ ,
\end{equation}
where $\sigma_1$ stands for the corresponding Pauli matrix. Using
well-known relation
\begin{equation}\label{}
  e^{i\pi/4\, \sigma_2}\, \sigma_1\, e^{- i\pi/4\, \sigma_2} =
  \sigma_3\ ,
\end{equation}
one obtains
\begin{eqnarray}\label{}
 e^{i\pi/4\, \sigma_2}\, S_2(z)\,  e^{- i\pi/4\, \sigma_2} &=&
 \exp\Big( z (\underline{a}^\dag)^T \sigma_3 \underline{a}^\dag   - z^*
  \underline{a}^T \sigma_3 \underline{a} \Big)\nonumber \\ &=&
  S^{(1)}(z) S^{(2)}(-z)\ ,
\end{eqnarray}
where $S^{(k)}$ denotes a single-mode $(k)$ squeeze operator. Now,
since $S^{(k)}(z)$ is unitarily equivalent to $S^{(k)}(r)$ a
two-mode squeeze operator $S_2(z)$ is unitarily equivalent to
$S^{(1)}(r)S^{(2)}(-r)$. Hence, the spectral properties of
$S_2(z)$ easily follows. In particular defining the space $\cal D$
of smooth functions $\psi=\psi(x_1,x_2)$ with compact supports and
${\cal Z} = F[{\cal D}]$ one obtains the following
representations:
\begin{eqnarray}  \label{S2-Z}  S^{(1)}(r)S^{(2)}(-r) =  \sum_{nm=0}^\infty
\, s^-_{nm}\, |f^+_{nm}\r \la f^-_{nm}|
\hspace{1cm} \mbox{on}\ \ \ {\cal Z} \ , \end{eqnarray} and
\begin{eqnarray} \label{S2-D} S^{(1)}(r)S^{(2)}(-r) =  \sum_{nm=0}^\infty
\, s^+_{nm}\, |f^-_{nm}\r \la
f^+_{nm}| \hspace{1cm} \mbox{on}\ \ \ {\cal D}\ ,
\end{eqnarray}
where
\begin{equation}\label{}
  f^\pm_{nm}(x_1,x_2) = f^\pm_n(x_1)f^\pm_m(x_2) \ ,
\end{equation}
and
\begin{equation}\label{s-nm}
  s^\pm_{nm} = e^{\pm r(n-m)}\ .
\end{equation}
Jannussis et al. \cite{series2} claimed that the eigenvalues of
$S_2(z)$ are given by $e^{2(m-n)}$, see e.g. formula (5.7) in
\cite{series2}. Their result has the similar form as $s^-_{nm}$
but of course it is incorrect. Note that eigenvalues of
\cite{series2} do not depend upon the squeezing parameter $z$ as
was already observed in \cite{Ma}.

\section{N-mode squeezing}

Following  \cite{Ma2} one defines an $N$-mode squeeze operator
\begin{equation}\label{SN}
  S_N( \widehat{Z}) = \exp\left( \frac 12 \,
  (\underline{a}^\dag)^T\,\widehat{Z}\,\underline{a}^\dag -
  \frac 12 \,
  \underline{a}^\dag\,\widehat{Z}^\dag\,\underline{a}\right)
\end{equation}
where $ \widehat{Z}$ is an $N\times N$ symmetric (complex) matrix
and
\begin{equation*}\label{}
  \underline{a}^T = (a_1,a_2,\ldots,a_N)\ .
\end{equation*}
Defining an $N$-mode rotation operator
\begin{equation}\label{}
  R_N( \widehat{\Phi}) = \exp\left( i ( \underline{a}^\dag)^T\,
  \widehat{\Phi}\, \underline{a} \right)\ ,
\end{equation}
with $ \widehat{\Phi}$ being an $N\times N$ hermitian matrix, one
shows \cite{Schumaker3,Ma2}
\begin{equation}\label{}
R_N^\dag( \widehat{\Phi})\, S_N( \widehat{Z})\, R_N(
\widehat{\Phi}) = S_N\left( e^{-i \widehat{\Phi}}\, \widehat{Z}\,
e^{-i \widehat{\Phi}^T}\right)\ .
\end{equation}
Now, by a suitable choice of $ \widehat{\Phi}$ one obtains
\begin{equation}\label{}
e^{-i \widehat{\Phi}}\, \widehat{Z}\, e^{-i \widehat{\Phi}^T} =
\widehat{Z}_{\rm D}\ ,
\end{equation}
where $ \widehat{Z}_{\rm D}$ is a diagonal matrix, i.e. $
(\widehat{Z}_{\rm D})_{kl} = z_k \delta_{kl}$. Hence, an $N$-mode
squeeze operator $S_N( \widehat{Z})$ is unitarily equivalent to
\begin{equation}\label{}
R_N^\dag( \widehat{\Phi})\, S_N( \widehat{Z})\, R_N(
\widehat{\Phi}) = S^{(1)}(z_1)\, S^{(2)}(z_2)\, \ldots \,
S^{(N)}(z_N)\ ,
\end{equation}
and therefore its properties are entirely governed by the
properties of the single-mode squeeze operator $S(z)$. In
particular $S_N( \widehat{Z})$ gives rise to a discrete family of
generalized eigenvalues being combinations of generalized
eigenvalues of $S^{(k)}(z_k)$. Defining the corresponding
subspaces $\cal D$ and $\cal Z$ in the Hilbert space $L^2(
\mathbb{R}^N)$ one easily finds
\begin{eqnarray}  \label{SN-Z} R_N^\dag( \widehat{\Phi})\, S_N( \widehat{Z})\, R_N(
\widehat{\Phi})  = \sum_{n_1, \ldots, n_N=0}^\infty \, s^-_{n_1
\ldots n_N}\, |f^+_{n_1 \ldots n_N}\r \la f^-_{n_1 \ldots n_N}|
\hspace{1cm} \mbox{on}\ \ \ {\cal Z} \ ,
\end{eqnarray} and
\begin{eqnarray} \label{SN-D}
R_N^\dag( \widehat{\Phi})\, S_N( \widehat{Z})\, R_N(
\widehat{\Phi}) =  \sum_{n_1, \ldots, n_N=0}^\infty \, s^+_{n_1
\ldots n_N}\, |f^-_{n_1 \ldots n_N}\r \la f^+_{n_1 \ldots n_N}|
\hspace{1cm} \mbox{on}\ \ \ {\cal D}\ ,
\end{eqnarray}
where
\begin{equation}\label{}
  f^\pm_{n_1 \ldots n_N}(x_1,\ldots,x_N) = f^\pm_{n_1}(x_1)\, \ldots f^\pm_{n_N}(x_N) \ ,
\end{equation}
and
\begin{equation}\label{sN-nm}
  s^\pm_{n_1 \ldots n_N} = \exp\left\{ \pm \left[ r_1\left(n_1 + \frac 12 \right) + \ldots
  + r_N \left( n_N + \frac 12\right)\right]  \right\}\ ,
\end{equation}
with $r_k = |z_k|$.

\section*{Acknowledgments}

 This work was partially
supported by the Polish Ministry of Scientific Research and
Information Technology under the  grant No PBZ-MIN-008/P03/2003.

\end{document}